# Enhancing superconductivity with resonant anti-shielding and topological plasmon-polarons

Krzysztof Kempa[1*], Nakib H. Protik[2], Tyler Dodge[1], Claudia Draxl[2] and Michael J. Naughton[1]

[1]Department of Physics, Boston College, Chestnut Hill, Massachusetts 02467, USA
[2]Institut für Physik and IRIS Adlershof, Humboldt-Universität zu Berlin, 12489 Berlin, Germany

## Abstract

By employing *ab initio* Migdal-Eliashberg calculations, we predict a 4-fold enhancement of the superconducting critical temperature of $MgB_2$ when proximity-coupled to the topological crystal $Bi_2Se_3$. We support this result with calculations using the general Leavens scaling method. We show that this effect is a result of *dynamic resonant anti-shielding* of Cooper pairs by plasmon polarons of Dirac electrons in the topological crystal. Our calculations show that such superconductivity enhancement varies strongly with Coulomb coupling between plasmon polarons and Cooper pairs, with a pronounced maximum of $T_c$ at a critical value of the coupling parameter. This feature is universal, and so can occur in other superconductor-topological crystal combinations, including with non-phonon mediated superconductors. We discuss methods to experimentally optimize the key coupling parameter.

Increasing the critical temperature $T_c$ of superconductivity toward room temperature has been a type of holy grail of physics. Discovery of the cuprate superconductors in the late 1980s, with $T_c$ up to 92 K, rekindled the field, and raised hopes that room temperature could be in sight. Cuprate $T_c$ was increased to 133 K [1] by 1993, but has since stalled there. Despite intense effort, similar sluggish progress has been made on the theoretical front, with the origin of cuprate superconductivity remaining insufficiently clear still today. Even though carrier bosonization remains a key concept, the pairing mechanism seems more subtle than the BCS electron-phonon-electron interaction. In an early paper, Ginzburg and Kirzhnits [2] argued that there are no physical limits to prevent room temperature operation of BCS superconductors.

Separately, following an early reformulation of BCS theory in terms of an effective dielectric function [2,3,4], metallic metamaterials have been proposed to increase $T_c$ by controlling the dielectric environment [5,6]. These can customize a dielectric response to obtain exotic optical properties [7], and were engineered to produce a small effective dielectric function, potentially capable of Cooper pair enhancement via anti-shielding. However, due to the locality of metamaterial dielectric response, only marginal $T_c$ enhancements were observed [5,6]. Here, we demonstrate theoretically that dramatic $T_c$ enhancements are possible via *resonant anti-shielding* (RAS), induced in the plasmon-polaron mode of the Dirac electrons on the surface of a topological crystal (TX) coupled to a superconductor, as sketched in Fig. 1.

To illustrate the basic physics of anti-shielding, we begin with the standard model of a dressed electron-electron interaction in a jellium metal [8], containing the basics of the Migdal-Eliashberg theory:

$$\frac{V_q}{\varepsilon} + \left|\frac{g_q}{\varepsilon}\right|^2 \frac{2\omega_q}{(\omega^2-\omega_q^2)+i\delta} = \frac{V_q}{\varepsilon_{eff}} \quad (1)$$

where $V_q = 4\pi e^2/q^2$ is the bare electrostatic potential, $\varepsilon$ the dielectric function of the environment, $g_q$ the matrix element for electron-phonon scattering, averaged over all electronic states, $\omega_q$ the phonon dispersion, and $i\delta$ a small constant loss factor. The first term in is the screened electron-electron interaction, and the second, Fröhlich term is the electron-electron interaction mediated by phonons with frequency $\omega_q$. Cooper pairing can occur at frequency $\omega \approx \omega_q$, with the wavevector $q$ of the order of $k_F$ (Fermi). Eq. (1) also shows that pairing can be strengthened by making $|\varepsilon| < 1$, *i.e.* anti-shielding, which represents enhancement, rather than suppression (shielding) of the interactions. RAS can occur if $|\varepsilon| \ll 1$. For a typical superconductor, $\varepsilon > 1$ and thus, anti-shielding is impossible without some additional mechanism. Similarly, conventional metamaterial structures cannot provide a RAS effect since, while vanishing of $\varepsilon$ at $\omega \approx \omega_q$ is possible, achieving this at $q \sim k_F$ is exceedingly difficult, as this would require the smallest structured feature sizes to be of the order of $1/k_F$ [9]. Only surface roughness could provide such a minute corrugation. Also, the screening is much stronger for the Fröhlich term in Eq. (1); any more realistic treatment would require spectral averaging which, as a result of $\varepsilon$ changing sign about the vanishing point, would lead to cancellations in the first term $\sim 1/\varepsilon$, and accumulations for the second, which goes as $1/|\varepsilon|^2$. The same holds for anti-shielding, including RAS.

Another strategy to achieve RAS lies in the fact that Maxwell's equations allow for the existence of longitudinal plasmon modes for which $\varepsilon(q, \omega) = 0$. However, conventional plasmon modes occur in a sector of phase space far from the required $\omega \approx \omega_q$ with $q$ of the order of $k_F$. Recently, an unusual plasmonic "$\alpha$–mode" was observed in the topological crystal $Bi_2Se_3$ [10], in that required phase space. The dispersion curve for this mode is close to linear, $\omega \propto q$ (Fig. 2a) and is clearly not a pure phonon mode, since it crosses the BZ edge without any momentum Umklapp [10]. The most striking observation was that this mode remains strong and extremely weakly damped, with damping rate and intensity almost constant for $2k_F < q < 6k_F$ [11]. All other known plasmon modes are unobservable in that range. An interesting observation was that in the non-topological form of $Bi_2Se_3$, this $\alpha$–mode disappears and is replaced by a conventional, transverse acoustic phonon mode (Fig. 2a). The new acoustic phonon mode has a standard dispersion, close to that of the $\alpha$–mode in the first BZ [10].

A recent theoretical study [11] is consistent with these discoveries. It shows that the $\alpha$–mode is a plasmon-polaron [12], a hybrid of plasmon excitations of Dirac surface electrons, and a transverse acoustic phonon mode. This $\alpha$–mode has topological character, with collective spin-charge fluctuations of the topological 2D Dirac band states at the surface. Ref. [11] demonstrated that the $\alpha$-mode has near perfect suppression of forward and backward scattering, resulting in ultralow damping, and an absence of Umklapp scattering at the BZ boundary. This $\alpha$–mode is similar to the phonon-polariton mode, a hybrid of photon and phonon excitations. To obtain the dispersion relation for the polariton [13], one starts with the dispersion relation for photons, $\omega = qc/\sqrt{\varepsilon}$, and replaces $\varepsilon$ with the Lyddane–Sachs–Teller phonon formula [14], $\varepsilon = \varepsilon_{eff} = \frac{\omega_{LO}^2 - \omega^2}{\omega_{TO}^2 - \omega^2}$. By analogy, one can derive the dispersion for the plasmon-polaron by starting with the dispersion for the topological 2D Dirac plasmon [11], which contains $\bar{V}_q = V_q/\varepsilon$. We assume that $\varepsilon = \varepsilon_{eff}$, except now $\varepsilon_{eff}$ is given by Eq. (1), with $\varepsilon = \bar{\varepsilon}$ (background dielectric constant). In the limit of interest in this work ($q \sim k_F$ and $\omega \sim \omega_q$), one gets

$$\varepsilon_{TI}(q,\omega) \approx 1 + \kappa |g_q|^2 \frac{2\omega_q}{(\omega^2 - \omega_q^2) + i\delta^2} \approx \frac{(\omega^2 - \bar{\omega}_q^2)}{(\omega^2 - \omega_q^2) + i\delta^2} \qquad (2)$$

with parameter $\kappa \sim k_F/\bar{\varepsilon}^2 > 0$, where $k_F$ is the Fermi wavevector of the surface electrons in Bi$_2$Se$_3$, and the plasmon-polaron frequency is given by $\bar{\omega}_q^2 \approx \omega_q^2 - 2\kappa |g_q|^2 \omega_q$, and so $\bar{\omega}_q < \omega_q$. This frequency is confirmed by experiment [11] and theory [12]: the plasmon-polaron mode is negatively depolarization-shifted, *i.e.* it follows the phonon mode in the first BZ, but always at frequencies lower than the phonon mode (see Fig. 2a).

Consider now a superconductor film sandwiched between two TX slabs, as sketched in Fig. 1. We assume that the superconductor is sufficiently thin ($t_{sup} < \zeta \leq 1/q$), so that RAS is uniformly extended throughout the superconductor. The topological proximity effect, discussed below, can significantly relax this requirement [15]. Then, the effective dielectric function experienced by electrons in the superconductor is given by $\bar{\varepsilon}_{sup}(q,\omega) \approx \varepsilon_{sup} + [\varepsilon_{TI}(q,\omega) - 1]$, where $\varepsilon_{sup}$ is $\varepsilon$ of the bulk superconductor, of order 1 in the required domain of phase space, and the term in the square parentheses is the polarizability of the Dirac surface electrons of the topological Bi$_2$Se$_3$.

Phonons of the superconductor control the behavior of the plasmon-polaron, and we generalize Eq. (2) by relaxing the jellium assumption and by including all relevant phonon bands. Then, with $\epsilon_\mathbf{k}$ the electron energy, $g_{\mathbf{kk}'\nu}$ the generalized matrix element for scattering between electronic states $\mathbf{k}$ and $\mathbf{k}'$ through a phonon with $\mathbf{q} = (\mathbf{k}' - \mathbf{k}, \omega_{\mathbf{q}\nu})$ in phonon branch $\nu$, and $\delta \to 0^+$, $\bar{\varepsilon}_{sup}$ becomes

$$\frac{\bar{\varepsilon}_{sup}(\omega)}{\varepsilon_{sup}} = \tilde{\varepsilon}_{sup}(\omega) = 1 - \kappa \left\{ \alpha^2 F(\omega) \ln \left| \frac{\omega_{max} - \omega}{\omega_{min} - \omega} \right| + \int_{\omega_{min}}^{\omega_{max}} \left[ \frac{\alpha^2 F(\bar{\omega}) - \alpha^2 F(\omega)}{\bar{\omega} - \omega} + \frac{\alpha^2 F(\omega)}{\bar{\omega} + \omega} \right] d\bar{\omega} + i\pi \alpha^2 F(\omega) \right\}. \qquad (3)$$

where we use the renormalized, dimensionless Eliashberg function $\alpha^2 F(\omega)$.

We can estimate the expected value of $\kappa$ from experiment, by considering $Bi_2Se_3$ interfacing vacuum, in which case the phonon spectrum of $Bi_2Se_3$ controls the physics of the plasmon-polaron. Here, we model the Eliashberg function with a single dominant peak as a rectangle of height $\alpha^2 F(\omega) = 1$ in the range $\omega_{min} < \omega < \omega_{max}$, and $\alpha^2 F(\omega) = 0$ otherwise. Then, using Eq. (3), we obtain the result in Fig. 2(b) by assuming that $\kappa = 1$. This is in quantitative agreement with the experimental result shown in Fig. 2(a) at $q = 0.53$ Å$^{-1}$, which represents the maximum observed frequency difference, with $\Delta = \frac{\omega_q - \bar{\omega}_q}{\omega_q} \approx 20\%$. Since $\kappa$ can be varied, we use it as an adjustable parameter in our present calculations. The Eliashberg function is screened, as is the generalized matrix element $|g_{\mathbf{kk'}\nu}|^2$, i.e. $\overline{\alpha^2 F(\omega)} = \alpha^2 F(\omega)/|\tilde{\varepsilon}_{sup}(\omega)|^2$. As mentioned, RAS occurs for $|\tilde{\varepsilon}_{sup}(\omega)| \ll 1$, and it strongly enhances the screened Eliashberg function. This is the main effect of RAS, and the next step is to calculate $T_c$ from this screening-renormalized Eliashberg function. We first employ the *ab initio* solver based on a direct solution of the coupled Eliashberg equations [16, 17]. To calculate $T_c$, we solve directly the Eliashberg equations in Ref. [18]. The electron-phonon coupling function $\lambda$ is computed from $\alpha^2 F$. Above the transition temperature, $\Delta$ vanishes. For the isotropic solver, we could choose to shield or to anti-shield $\alpha^2 F$. In principle, the same can be done for the anisotropic case. However, while only the anisotropic theory correctly predicts the observed two-gap superconductivity in $MgB_2$, it is also known to *overestimate $T_c$* in the absence of screening. Since the isotropic solver *underestimates $T_c$* by about the same fraction, we conservatively chose this solver for studying anti-shielding, and the fully anisotropic solver only to validate our code. Thus, our calculation with screening/RAS is expected to lead also to an underestimation of $T_c$, and consequently the ratio of $T_c$, with and without screening, is a rational way to quantify the superconductivity enhancement. Further details can be found in the Supplemental Material.

To support our *ab initio* calculations, we apply also the Leavens scaling method [19]. In contrast to many others (see [16]), this method is valid (as is *ab initio*) for arbitrary strength $\lambda = \int_0^\infty \frac{\alpha^2 F(\omega)}{\omega} d\omega$, required while dealing with RAS. The scaling method estimates not $T_c$, but its upper limit, i.e. $T_c^{max} = c(\mu^*) \int_0^\infty \alpha^2 F(\omega)d\omega$. The term $c(x)$ is a monotonically decreasing function of $x$ (see Ref. 19) and the Coulomb pseudopotential is

$$\mu^* \approx \frac{N(\mu)U}{1+N(\mu)U \ln\left(\frac{\epsilon_F}{\hbar\overline{\omega}_q}\right)} \quad (4)$$

where $\epsilon_F$ is the Fermi energy, and $U$ is the double Fermi surface average of the screened Coulomb potential. Here, $\ln\left(\frac{\epsilon_F}{\hbar\overline{\omega}_q}\right)$ typically ranges from 5 to 10, and $N(\mu)U \gg \mu^*$. Thus, one can approximate Eq. (4) with $\overline{\mu^*} \approx \mu^* \approx \frac{1}{\ln\left(\frac{\epsilon_F}{\hbar\overline{\omega}_q}\right)}$, i.e. independent of $\tilde{\varepsilon}_{sup}(\omega)$. The general formula for $T_c^{max}$ including anti-shielding, is given in this method by

$$T_c^{max} = c(\mu^*) \int_0^\infty \overline{\alpha^2 F(\omega)} d\omega = c(\mu^*) \int_0^\infty \frac{\alpha^2 F(\omega)}{|\tilde{\varepsilon}_{sup}(\omega)|^2} d\omega = \frac{c(\mu^*)}{\kappa\pi} \int_0^\infty \text{Im}\left(\frac{1}{\tilde{\varepsilon}_{sup}(\omega)}\right) d\omega. \quad (5)$$

We apply both calculational methods to MgB$_2$, the acknowledged highest $T_c$ BCS-type superconductor (at ambient pressure), interfaced with Bi$_2$Se$_3$. We assume that phonons of MgB$_2$ control also the plasmon-polaron, and employ the *ab initio*-calculated $\alpha^2 F(\omega)$, with $\mu^* = 0.16$ [18,19]. Fig. 3(a) shows the resulting $T_c$ versus $\kappa$ (red-solid circles), calculated by *ab initio* solving the Eliashberg equations, and assuming a uniform field of the plasmon-polaron Bi$_2$Se$_3$ inside MgB$_2$ (sufficiently thin film of MgB$_2$). For $\kappa = 0$ (absence of screening), the calculated $T_c$ of 23 K is substantially lower than the experimental result of $T_c = 39$ K. The anisotropic calculation yields $T_c = 54$ K (open circle), *i.e.* substantially larger than experiment. $T_c$ steadily increases with increasing $\kappa$, for $\kappa = 1.3$ has its maximum of ~100 K, and after that, rapidly decays. At the maximum, there is about 4-fold enhancement of $T_c$, as compared to the case without screening ($\kappa = 0$).

Fig. 3(a) shows also the $T_c^{max}$ versus $\kappa$ result (solid lines) obtained from Eq. (5), *i.e.* employing Leavens scaling. Surprisingly, $T_c^{max} = 43$ K at $\kappa = 0$, is much closer to the experimental result than the *ab initio* result, but the overall, qualitative shapes of all the scaling curves are the same. In fact, these curves are quite close to the *ab initio* result, with a significant departure only at the critical $\kappa = 1.3$ (see Fig. 3a). This divergence results from the fact that at the RAS condition, $\tilde{\varepsilon}_{sup}(\omega)$ nearly vanishes, which can lead to a near singular behavior of the screened Eliashberg function. While the *ab initio* calculations seem unaffected by the problem, the scaling is affected. To remedy this, a small residual imaginary contribution $i\zeta$ (*e.g.* impurity scattering) can be added to $\tilde{\varepsilon}_{sup}(\omega)$, given by Eq. (3). The solid lines in Fig. 3(a) are calculated by varying $\zeta$. Clearly, the main effect of this correction is to soften the divergence, and to drive the $T_c^{max}$ curves closer to the *ab initio* result. The overall qualitative behavior of the curves in Fig. 3 can be understood analytically by using a toy model which employs $\varepsilon_{TI}(q,\omega)$ given by Eq. (2) as $\tilde{\varepsilon}_{sup}(\omega)$ in the very last part of Eq. (5). The resulting approximate formula is $T_c^{max} \sim \left[\left(\omega_q^2 - \kappa\right)^2 + \delta^4\right]^{-1/4}$. For $\kappa = 0$, it gives a finite result, at $\kappa = \omega_q^2$ it reaches a sharp maximum, and for $\kappa \to \infty$, it vanishes. These are the characteristics of all curves shown in Fig. 3.

Analysis of the above calculations indicates that further increasing of $T_c$ by RAS is possible, if the superconductor phonon spectrum is not simultaneously controlling the Cooper pairing and the plasmon polaron. This dominance of MgB$_2$ phonons forces (by the Kramers-Kronig relations) the near vanishing of $\tilde{\varepsilon}_{sup}(\omega)$ into the spectral domain of the nearly vanishing Eliashberg function. One intriguing possibility to avoid that would be to sandwich a TX (*e.g.* Bi$_2$Se$_3$) with a non-phonon mediated superconductor (*e.g.* YBCO) on one side, and a phonon robust material (*e.g.* MgB$_2$) on the other. We consider such a case by: (a) using an experimentally-retrieved Eliashberg function for YBa$_2$Cu$_3$O$_{7-\delta}$ (YBCO) [20] with $c(\mu^*) = 0.2$; (b) by calculating the dielectric response from the *ab initio*-calculated Eliashberg function for MgB$_2$ (using Eq. (3)); and (c) by applying Leavens scaling. Surprisingly again, for $\kappa = 0$, $T_c^{max} \approx 80$ K, close to the 92 K experimental value for YBCO, even though $c(\mu^*) = 0.17$ was chosen from the typical BCS range. As expected, the scaling method predicts diverging $T_c^{max}$ at the critical point (again, near $\kappa = 1.3$), and to quench it, we use (as before) $i\zeta$. Fig. 3(b) shows $T_c^{max}$ *vs.* $\kappa$ for this case, with each line calculated for a different $\zeta$. The $T_c^{max}$ divergence, clearly visible on the curve for $\zeta = 0.01$, is strongly damped for $\zeta = 0.05$, and completely disappears for $\zeta \geq 0.1$. By analogy to Fig. 3(a), one might expect the line for $\zeta = 0.1$ is not far from the $T_c$ *vs.* $\kappa$ line and therefore, (based on Fig. 3b) this structure might provide superconductivity at $T_c > 300$ K.

The dimensionless coupling parameter $\kappa$, which controls the $T_c$ enhancement, is given by $\kappa = \beta k_F/\bar{\varepsilon}^2$, with $k_F$ the Fermi wavevector of the plasmon-polaron host, $\bar{\varepsilon}$ the background dielectric constant of the plasmon-polaron host, and $\beta$ a constant that depends on the experimental superlattice geometry. Thus, $\kappa$ can be experimentally controlled, *e.g.* through $k_F$ by adjusting the doping level in the plasmon-polaron host, and/or through $\bar{\varepsilon}$, by changing the superlattice. In addition, such $T_c$ engineering might benefit from the topological proximity effect [15], at the surface of the TX TlBiSe$_3$ coated with superconducting Pb. It was shown that the topological state of the crystal extends up to 20 monolayers into the superconductor, without any admixing. This effect is expected to improve the plasmon-polaron penetration into the superconductor films of the superlattice, as well as could increase the efficiency of the phonon modifier layers.

Further possible architectures include natural or engineered bulk SC-TX layered materials, wherein the properties of the superconductor are modulated by the properties of the proximate TX. For example, the cuprates consist of hole- or electron-doped CuO$_2$ layers sandwiched by nonconducting layers (*e.g.* yttrium- or bismuth- oxide). One could consider synthesizing cuprate systems modified to incorporate known TX layers (*e.g.* chalcogenides). Similarly, many organic superconductors are comprised of 2D superconducting layers sandwiched by nonconducting layers, the latter of which might be engineered to have TX character. The same *in situ* strategy could be applied to MgB$_2$, a BCS superconductor with very large, relevant phonon frequencies. Such incorporated topological modifications could produce atomic/molecular layers functioning as charge reservoirs as well as providing the $T_c$-enhancing RAS effect. These kinds of systems could facilitate high temperature superconductivity in multiple physical forms, from single crystalline to nanocrystalline / ceramic, so long as the core TX–superconductor–TX character was preserved.

In conclusion, we have demonstrated that a plasmon-polaron residing at the surface of a topological crystal interfaced with a superconductor can resonantly anti-shield Cooper pairs in the superconductor. This anti-shielding occurs regardless of the pairing mechanism, and leads to multi-fold enhancement of $T_c$.

* Corresponding author: kempa@bc.edu

**Acknowledgements**

NHP was supported by a Humboldt Research Fellowship from the Alexander von Humboldt Foundation, Bonn, Germany.

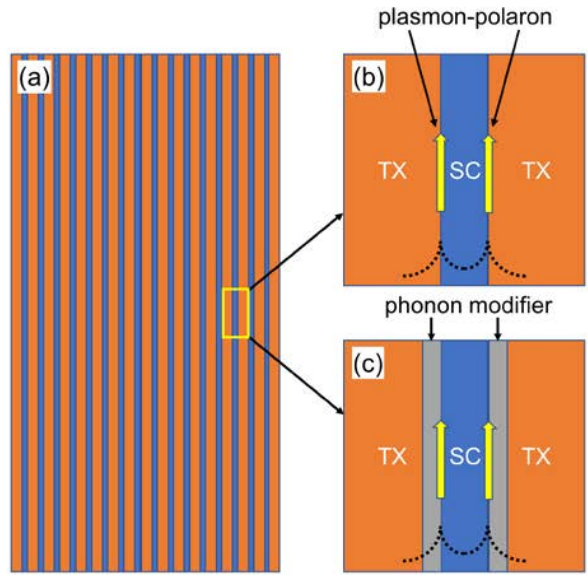

**Fig. 1.** (a) Superconductor (SC) -topological crystal (TX) superlattice structure designed to exploit the proposed resonant anti-shielding (RAS) effect produced by a surface plasmon-polaron. (b) Expanded view of the superlattice, also indicating the decaying amplitudes of the electric field (dashed lines) produced by the plasmon-polaron mode propagating (yellow arrows) along each interface. (c) Alternate structure containing additional phonon-modifier films. Note that, due to the topological proximity effect, plasmon-polaron modes occur at the interfaces of the modifiers with the superconductor.

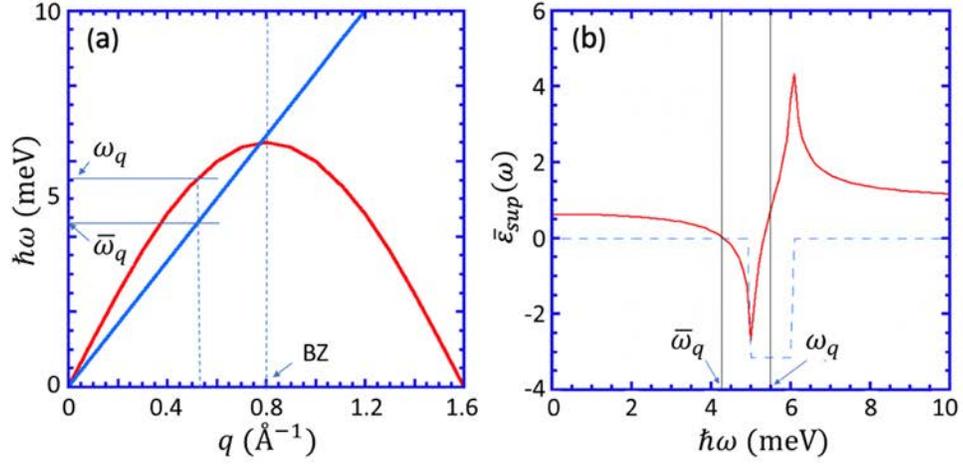

**Fig. 2.** (a) Collective modes of the 2D Dirac electron gas on the surface of a topological crystal $Be_2Se_3$ (interpolated from experimental data of Ref. [10]): $\alpha$–mode (blue line), acoustic phonon mode (red line). BZ - Brillouin zone. (b) Calculated dielectric function of the 2D Dirac electron gas using Eq. (3) with a step model of $\alpha^2 F$, and assuming $\kappa = 1$. Solid line: real part, dashed line: imaginary part.

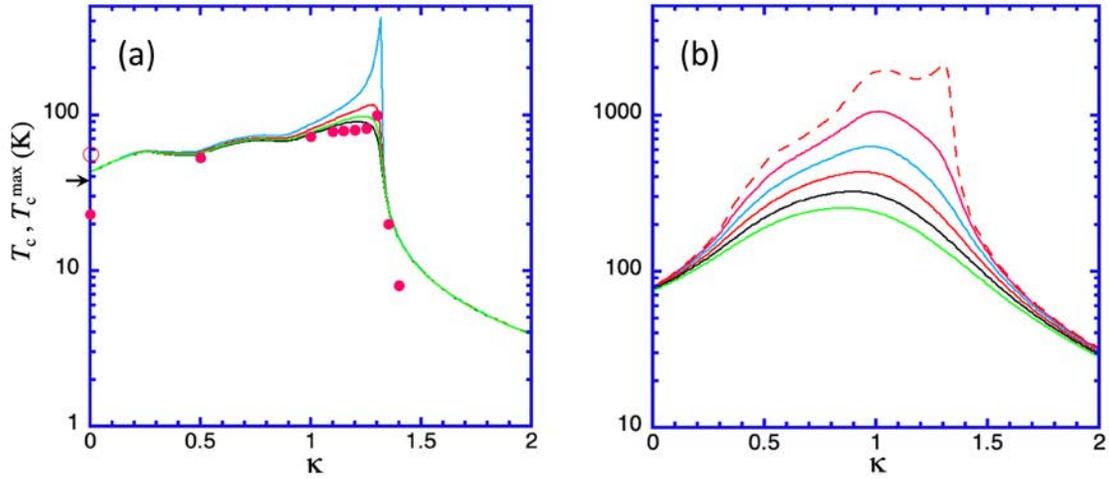

**Fig. 3** (a) Calculated superconducting critical temperature $T_c$ vs. coupling parameter $\kappa$ using the isotropic *ab initio* Eliashberg equations solution method for $MgB_2$, proximity coupled to $Bi_2Se_3$ (solid circles). Lines represent $T_c^{max}$ vs. $\kappa$ obtained from the Leavens scaling method, for different quenching parameters $\zeta$: 0 (blue line), 0.005 (red line), 0.008 (green line), and 0.01 (black-line). Open circle at $\kappa =0$ is for the anisotropic case (see text). Arrow indicates experimental $T_c$. (b) $T_c^{max}$ vs. $\kappa$ obtained from Leavens scaling for the YBCO- $Bi_2Se_3$ -$MgB_2$ structure, for different quenching parameters $\zeta = 0.01$ (dashed-red line), 0.05 (purple line), 0.1 (blue line), 0.15 (red line), 0.2 (black line), and 0.25 (green line).

**References**

[1] A. Schilling, M. Cantoni, J.D. Guo, and H.R. Ott, "Superconductivity above 130 K in the Hg–Ba–Ca–Cu–O system", *Nature* **363**, 56-58 (1993). doi: 10.1038/363056a0.

[2] V.L. Ginzburg and D.A. Kirzhnits, "On the problem of high temperature superconductivity", *Phys. Rep.* **4**, 343-356 (1972). doi: 10.1016/0370-1573(72)90017-8.

[3] D.A. Kirzhnits, E.G. Maksimov, and D.I. Khomskii, "The description of superconductivity in terms of dielectric response function", *J. Low Temp. Phys.* **10**, 79-93 (1973). doi: 10.1007/BF00655243

[4] V.L. Ginzburg, "The problem of high-temperature superconductivity II", *Sov. Phys. Usp.* **13**, 335-352 (1970). doi: 10.1070/PU1970v013n03ABEH004256.

[5] I.I. Smolyaninov and V.N. Smolyaninova, "Theoretical modeling of critical temperature increase in metamaterial superconductors", *Phys. Rev. B* **93**, 184510 (2016) doi: 10.1103/PhysRevB.93.184510; V. Smolyaninova, C. Jensen, W. Zimmerman, J.C. Prestigiacomo, M.S. Osofsky, H. Kim, N. Bassim, Z. Xing, M.M. Qazilbash, I.I. Smolyaninov, "Enhanced superconductivity in aluminum-based hyperbolic metamaterials", *Sci. Rep.* **6**, 34140 (2016). doi: 10.1038/srep34140.

[6] I.I. Smolyaninov and V.N. Smolyaninova, "Metamaterial superconductors", *Nanophotonics* **7**, 795-818 (2018). doi: 10.1103/PhysRevB.91.094501.

[7] D.R. Smith, J.B. Pendry, and M.C.K. Wiltshire, "Metamaterials and negative refractive index", *Science* **305**, 788-792 (2004). doi: 10.1126/science.1096796.8.

[8] R.D. Mattuck, "A guide to Feynman diagrams in the many-body problem", *Dover Publications, Inc.*, New York, 1976.

[9] A.J. Shvonski, J. Kong, and K. Kempa, "Nonlocal extensions of the electromagnetic response of plasmonic and metamaterial structures", *Phys. Rev. B* **95**, 045149 (2017). doi: 10.1103/PhysRevB.95.045149.

[10] X. Jia, S. Zhang, R. Sankar, F-C Chou, W. Wang, K. Kempa, E.W. Plummer, J. Zhang, X. Zhu, J. Guo "Anomalous acoustic plasmon mode from topologically protected states", *Phys. Rev. Lett.* **119**, 136805 (2017). doi: 10.1103/PhysRevLett.119.136805.

[11] A. Shvonski, J. Kong, K. Kempa, "Plasmon-polaron of the topological metallic surface states", *Phys. Rev. B* **99**, 125148 (2019). doi: 10.1103/PhysRevB.99.125148.

[12] I. Bozovic, "Low-energy collective electronic excitations in a polaron gas", *Phys. Rev. B* **48**, 876-880 (1993). doi: 10.1103/PhysRevB.48.876.

[13] G. Burns, "*Solid State Physics*", Academic Press, Inc., Orlando, 1985.

[14] R. Lyddane, R. Sachs, E. Teller, "On the polar vibrations of alkali halides", *Phys. Rev.* **59**, 673-676 (1941). doi: 10.1103/PhysRev.59.673.

[15] C.X. Trang, N. Shimamura, K. Nakayama, S. Souma, K. Sugawara, I. Watanabe, K. Yamauchi, T. Oguchi, K. Segawa, T. Takahashi, Y. Ando, T. Sato, "Conversion of a conventional superconductor into a topological superconductor by topological proximity effect", *Nat. Comm.* **11**, 159 (2020). doi: 10.1038/s41467-019-13946-0.

[16] E.F. Marsiglio and J.P. Carbotte, "Electron-phonon superconductivity", in K.H. Bennemann, J.B. Ketterson, (eds.) *Superconductivity* (Springer, Berlin, Heidelberg), pp. 73-162 (2008). doi: 10.1007/978-3-540-73253-2_3; F. Marsigilio, "Eliashberg theory: A short review", *Ann. Phys.* **417**, 168102 (2020). doi: 10.1016/j.aop.2020.168102.

[17] G.M. Eliashberg, *Zh. Eksperim. i Teor. Fiz.* **38**, 966-972 (1960) ["Interactions between electrons and lattice vibrations in a superconductor", *Soviet Phys. JETP* **11**, 696-702 (1960)].

[18] E.R. Margine, E. Roxana, and F. Giustino. "Anisotropic Migdal-Eliashberg theory using Wannier functions", *Phys. Rev. B* **87**, 024505 (2013). doi: 10.1103/PhysRevB.87.024505.

[19] C.R. Leavens, "A least upper bound on the superconducting transition temperature", *Solid State Commun*. **17**, 1499-1504 (1975). doi: 10.1016/0038-1098(75)90982-5.

[20] E.G. Maksimov, M.L. Kuli, O.V. Dolgov, "Bosonic spectral function and the electron-phonon interaction in HTSC cuprates", *Adv. Condens. Matter Phys.* **2010**, 423725 (2010). doi: 10.1155/2010/423725.